\documentclass[preprint2]{aastex}

\shorttitle{ZAMS Central Temperature}
\shortauthors{Rowe, Richer, Brewer, Crabtree}

\begin{document}

\title{Carbon Stars and other Luminous Stellar Populations in M33}
\author{J.F. Rowe\footnote{Visiting Astronomer, Canada-France-Hawaii Telescope (CFHT), operated by the National Research Council of Canada, le Centre National de la Recherche Scientifique de France, and the University of Hawaii}, H.B. Richer, J.P. Brewer$^1$\\
University of British Columbia \\
6224 Agricultural Road, Vancouver, BC, V6T 1Z1\\
rowe@astro.ubc.ca,richer@astro.ubc.ca,jbrewer@bcit.ca\\
\ \\
D.R. Crabtree$^1$\\
Dominion Astrophysical Observatory, Herzberg Institute of Astrophysics, National Research Council\\
5071 W. Saanich Road, Victoria, BC, V8X 4M6\\
dennis.crabtree@nrc.ca\\
\vspace{-.4in}
}
\begin{abstract}

The M33 galaxy is a nearby, relatively metal-poor, late-type spiral.
Its proximity and almost face-on inclination means that it projects
over a large area on the 
sky, making it an ideal candidate for wide-field CCD mosaic imaging.
Photometry was obtained for more than $10^6$ stars covering a 
$74\arcmin \times 56\arcmin$ field
centered on M33. Main sequence (MS), supergiant branch (SGB), red giant
branch (RGB) and asymptotic giant branch (AGB) populations
are identified and classified based on broad-band $V$ and $I$
photometry.  Narrow-band filters are used to measure spectral features
allowing the
AGB population to be further divided into C and M-star types.  The
galactic structure of M33 is examined using
star counts, colour-colour and colour-magnitude selected stellar
populations.  We use the C to M-star ratio to investigate the metallicity
gradient in the disk of M33.  The C/M-star ratio is
found to increase and then flatten with increasing galactocentric radius in
agreement with viscous disk formation models.  The C-star luminosity
function is found to be similar to M31 and the SMC, suggesting that
C-stars should be useful distance indicators. The ``spectacular arcs
of carbon stars'' in M33 postulated recently by \citet{blo04} are
found in our work to be simply an extension of M33's disk.

\end{abstract}

\vspace{0.2in}

\section{Introduction} 

 Carbon-star (C-star) production is caused when deep stellar
convection dredges up material created by nuclear fusion processes.  
Whether a star is a C-star or M-star depends on the C/O ratio in
its photosphere. If a star is formed from initially
metal-poor material in the original protostellar cloud then
less carbon is required to alter the surface chemistry from
oxygen dominated (C/O $<\ $ 1) to carbon dominated (C/O $>\ $ 1).
Photospheric chemistry is dominated by the production of CO
molecules.  An over-abundance of carbon leads to molecules such as CN
being formed, whereas an over-abundance of oxygen leads to production
of molecules such as TiO.  An AGB star initially has an
oxygen dominated photosphere, inhibiting the formation of CN as any
carbon primarily forms CO.  When a star undergoes dredge up, carbon rich
material mixes into the photosphere.  If
there is initially a low oxygen abundance in
the star then little carbon is needed to transform it from an M-star
into a C-star. 
Thus, the ratio of the number of C-stars to M-stars will depend on the
initial metallicity of the system, and observations support the idea that
higher C/M ratios occur in lower metallicity
systems and galaxies (\citet{bla83}; \citet{ric85a}; \citet{coo86};
\citet{mou86}; \citet{aar87}; \citet{bre95}; \citet{alb00}).  The
observed correlation spans 4 dex in C/M and 1.5 dex in [Fe/H]
\citep{gro02} and holds regardless of the galaxy morphology or star 
formation history.  This provides a method to
measure the metallicity distribution within a galaxy as only age and
metallicity appear to have a strong effect on the C/M-star ratio.      

\begin{figure}[htbp]
\begin{center}
\leavevmode
\includegraphics[width=7cm,height=4.42cm]{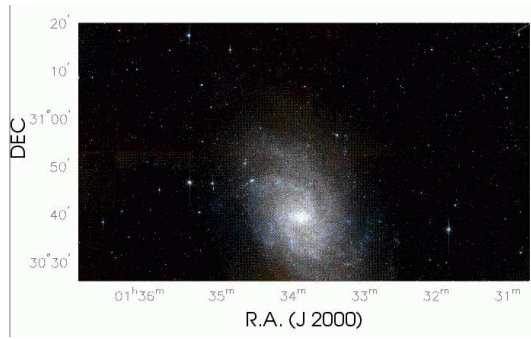}
\end{center}
\caption{M33 mosaic V-band image constructed from the CFHT images and covering the entire area of the present survey.}
\label{fig:m33V}
\end{figure}

Located in the Triangulum constellation, M33, also known as the
Triangulum Galaxy, is a late-type spiral located  
approximately 840 kpc away.  A V-band image using data from this work
is shown in Figure \ref{fig:m33V}.  M33 is substanially smaller in
size and mass than both M31 and 
the Milky Way and is an interesting target to study as its 
AGB/RGB stellar content is easily resolved in 4m class telescopes.
In contrast to M31, M33's lack of nearby dwarf companions provides 
it with an almost isolated  environment.

AGB stars are members of intermediate aged (1-10
Gyr) stellar populations and represent a relaxed
subsystem in galaxies \citep{now01}.  This means that AGB stars
uniquely record the star formation history of the galaxy at intermediate ages
as well as sampling its history of minor mergers.  
By observing M33 to large galactocentric
distances we can examine the underlying stellar population to see if
there is evidence for recent tidal
interactions. For example the newly discovered tidal ring that
appears to surround the Milky Way was identified through observations
of an F-star overabundance (\citet{new02}; \citet{iba03}).  

Using the C/M-star ratio one can trace metallicity variations in a
galaxy.  Zaritsky's \citep{zar92} 
star-forming viscous disk models predict a change in the slope of the
metallicity gradient at the radius where the rotation curve
flattens.  The current data set allows us to measure the metallicity
gradient of M33 as a function of galactocentric radius, thus
allowing tests of galaxy formation
and evolution models to be made.   

\subsection{C-star Classification}

In order to distinguish C and M-type AGB stars, groups led by
Richer (\citet{ric84}; \citet{ric85a}; \citet{ric85b}; \citet{pri87};
\citet{hud89}; \citet{ric90}) and Aaronson (\citet{aar84},
\citet{coo86}) developed a four band photometric system (FBPS) to
classify AGB stars.

The FBPS uses two
narrow-band filters to provide low-resolution spectral information and
two broad-band filters for temperature information.  The filters used are
listed in Table \ref{ta:filters}.  A C-star spectrum will
have CN bands, whereas an M-star is dominated by
oxide bands such as H$_2$O and TiO.  The
CN and TiO filters were developed to measure the CN and
TiO molecular band strengths.  Figure 11 of \citet{bre96} illustrates
spectra of a C-star, an M-star and an A-star and demonstrates how the
filters easily discriminate between the three.  A C-star will have
strong absorption in the 
CN filter and when compared to the magnitude measured in TiO
will produce a positive CN$-$TiO index.  An
M-star will produce a negative CN$-$TiO index, as it
will exhibit strong absorption from TiO. An A-star will produce a
CN$-$TiO index of approximately zero.  In the study of \citet{bre96}
the validity of the system was confirmed by spectroscopic observations.

The FBPS allows large areas to be quickly surveyed by
direct imaging, providing simultaneous measurements of all stars in the
field of view. This is in contrast to
spectroscopic observations which are limited to a relatively narrow
field of view, small numbers of potential targets and longer
integration times.  The FBPS is also advantageous as it will work in
fields too crowded for grisms. Spectroscopic survey strategies applied
to the LMC (\citet{bla80}, \citet{bla83}) would not work in the
present case on account of the faintness and crowding of M33's stars. 
A spectroscopic survey of all potential
AGB stars in M33 for the purpose of classification is
unfeasible.  Using the FBPS, stars can be quickly classified and
targeted for follow up spectroscopic studies. 

\section{Observations}

Multiband photometric data were collected on October 30 and 31, 1999
and December 3 and 4, 2000, with the 3.58 m
Canada-France-Hawaii Telescope (CFHT).
The detector used was the CFH12k mosaic CCD camera which employs 12
MIT/LL CCID20 CCDs to provide an effective size of 12 228$\times$8192
pixels.  The camera is positioned at prime focus, has a pixel size of
15 microns, a plate scale of 0$\farcs$206 /pixel and
42\arcmin$\times$28\arcmin \ field of view (approximately 1.5
times the size of the full Moon).  

The centres of the four fields in M33 that were observed are listed in Table
\ref{ta:target}.  The total observed area covered by the 4 fields is
80\arcmin$\times$50\arcmin .  An observing log is given in Table
\ref{ta:obslog}. 

\section{Data Reduction}

The science images required correction of
bad pixels, overscan and bias subtraction and flat-fielding.  These
operations were completed using the MSCRED package in IRAF\footnote{Image
Reduction and Analysis Facility (IRAF), a software package distributed
by the National Optical Astronomy Observatories (NOAO)}.

Much of the CFH12k is free of defects, such as bad columns and 
hot pixels, but some CCDs show significant cosmetic flaws, such as
CCD05 where 
approximately 30\% of the CCD pixels are defective.  Defective regions
often show a non-linear  
response to the number of incident photons.   For CFH12k, the number
of ADUs per pixel at which non-linearity 
becomes significant is different for each CCD and ranges from 51k to
65k.   Bad pixels
and columns can be treated in two ways.  The first is to ignore them
in the reduction, the second is to interpolate over these pixels using 
surrounding pixels.  The general approach taken in this work is to
simply ignore bad pixels, especially since the images are dithered so
chances are good that every part of the target will be observed at
least once.  Only in very specific cases
do we apply a correction to bad pixels, which we will describe in
\S\ref{broadpro}.  The broad-band data (obtained October, 1999)
and narrow-band data (obtained December, 2000) were
reduced in different ways as described below. 

\subsection{Narrow-band Data Processing}

After bias corrections, gain differences across the CCD were corrected
for by combining twilight flats for each filter.  When combining
frames, a 3 sigma clipping criterion (in MSCRED.COMBINE) is used to 
eliminate high and low pixel values.  This works well for the sharp bright
centres of stars' PSFs, but the extended wings are too faint to be
excluded.  To overcome this problem, a star's PSF is used as a
tracer to reject all pixels within a specified radius.  For
CFH12k, a radius of 15 pixels was found to work well from visual
examination of the images.

\subsection{Broad-band Data Processing} \label{broadpro}

The flat-fields for the broad-band data set suffer from a strong,
variable, scattered light pattern.  Flat-fielding with these images
introduced a 20\% 
response error as a smooth gradient.  The scattered
light signal was also found to dramatically change with each
flat-field image.  This effect meant that when the images were averaged
together, sigma clip routines to remove stars failed as it was impossible to 
scale each image
to a uniform level so that deviant pixel values could be reliably removed.  
To fix this problem
each flat-field image was heavily smoothed with a 500$\times$500 pixel 
mean boxcar filter leaving behind the slowly varying background.
Subtracting this signal away from the original flat-field image leaves an image
containing the pixel-to-pixel sensitivity changes and objects such as
stars and cosmic rays.  A bad
pixel mask was used to identify and correct the defects with 
interpolation from nearby pixels. 

The residual images from subtracting the smoothed flat-field images
were averaged using a sigma clipping algorithm to reject
pixels affected by stars or cosmic rays.
This flat-field corrects pixel-to-pixel sensitivity differences but failed to
remove the overall gradient.  This instrumental signature was removed
by using the science images
themselves.  It was assumed that the individual CCD fields
on the mosaic, which were farthest from the centre of the galaxy and
hence least contaminated by stars, should be completely flat.  The
assumption of intrinsic flatness is justified as according to the NASA
Extragalactic Database (NED) M33 reaches 25 
mag arcsec$^{-2}$ in the B-filter at a major axis radius of 70$\farcm$8.
The ratio of the 
major axis to the minor axis is 1.70.  The average B$-$V index of M33
reported by NED is 0.55 and the sky at Mauna Kea is approximately
V=21.7 mag arcsec$^{-2}$ \citep{kri90}. 
Individual CCD images that are outside an ellipse centred on M33 with
a major axis radius of 
70$\farcm$ and the same ellipticity as M33 were averaged to create a
{\it super flat} which was then 
normalized to unity.  Stars were removed from the
individual images using PSF fits and also applying the same sigma clipping
algorithm used for the creation of regular flat-fields.  A minimum of
3 images per CCD were  combined to 
create the super flat-field.  Since this calibration is being used for
the removal of a slowly changing gradient, the image was smoothed
using a 3$\times$3 mean boxcar.  Subtraction of the smoothed image from the
original showed a flat image consistent with the expected noise
level. The flat-fielded science images were then divided by the super
flat-field to remove the instrumental gradient.  

\subsection{Photometry} \label{photometry}

All stellar photometry was performed using the DAOPhot/ALLSTAR
package (\citet{ste87}, \citet{ste94}).  As DAOPhot is unable to
handle multi-extension FITS images each mosaic frame was split into its
individual frames.  This gave a total of 612 images which were all treated
independently for the extraction of photometric data. 

In order to account for geometric distortions in the PSFs, 150
stars were selected on each image and used to construct a
quadratically varying PSF.  The 
photometric measurements were made with ALLSTAR, which
simultaneously fits groups of stars found close to each other on the
frame with the PSF. 
Using photometry from each CCD chip, registration of the chips
relative to one another was done with DAOMatch/DAOMaster.  This worked
well with each individual chip and a 20 parameter transformation 
was used to 
model the geometric distortions, so that the measured pixel positions
of the stars on each CCD chip could be matched to other observations of
the same field on the same chip.  For example, for CCD01 in Field-2,
there were 3 sets of observations in each of the four filters.  These
12 images were then registered to match common objects for each
field.  Each observation was slightly offset from one another, meaning
that some stars were observed on two adjacent detectors.  Using
these common observations, the complete photometric catalogue was
pieced together, placing all objects on a common co-ordinate system as
described in the next section.

\subsection{Astrometric Registration} \label{registration}

In order to identify common objects between adjacent CCDs the pixel
co-ordinate system had to be transferred to
J2000 co-ordinates.  The rationale for this procedure is that 
DAOMatch/DAOMaster failed to 
converge to a proper registration solution, as less than 5\% of the
area of two chips overlapped from the dithered observations.  When
DAOMaster finds a transformation, it is only valid for the objects in
common between the images.  Extrapolation of the solution to adjacent
CCDs would introduce large distortions.  Instead, the pixel co-ordinate
system for each CCD chip was first mapped to the J2000 co-ordinate system.

Common stars between this survey and the USNO-A2 astrometric catalogue
were identified by using the MSCZERO and CCFIND IRAF commands found in
the MSCRED and IMCOO packages.  
The CCMAP program was used to automatically cross-identify 100
common stars in each CCD frame by finding the brightest star within
a 20$\times$20-pixel search box.  The success rate was approximately 75\%,
with a majority of failures due to catalogue stars located outside the
imaged area.  The
CCMAP program was then used to compute a rough astrometric solution
based on all cross-identifications, including incorrect ones, since
the number of
true matches dominates the list.  These solutions had an RMS error of
approximately 6\arcsec.  With this plate solution, the Starlink
package {\sc GAIA} was used to identify common stars between the two
catalogues. The output from {\sc GAIA} was input into CCMAP
to compute an accurate plate solution.  The average RMS error
was reduced to 0$\farcs$5 or approximately 2 pixels, which is about the
internal accuracy of the USNO-A2 catalog.  The entire photometric
catalogue was then transformed onto the J2000 co-ordinate system.
  
This new catalogue was searched for duplicate objects that were
imaged on adjacent CCDs.  These objects were located by identifying the
closest neighbour to each object and the closest objects to that
neighbouring star.  If two stars were found to be closer than
1.5\arcsec \ and their instrumental magnitudes differed by less than
0.1 magnitudes, then those stars were assumed to be the same and
combined into a single entry.

\subsection{Photometric Calibration} \label{photcal}

\begin{figure}[htbp]
\begin{center}
\leavevmode
\includegraphics[width=7cm,height=9.60cm,angle=270]{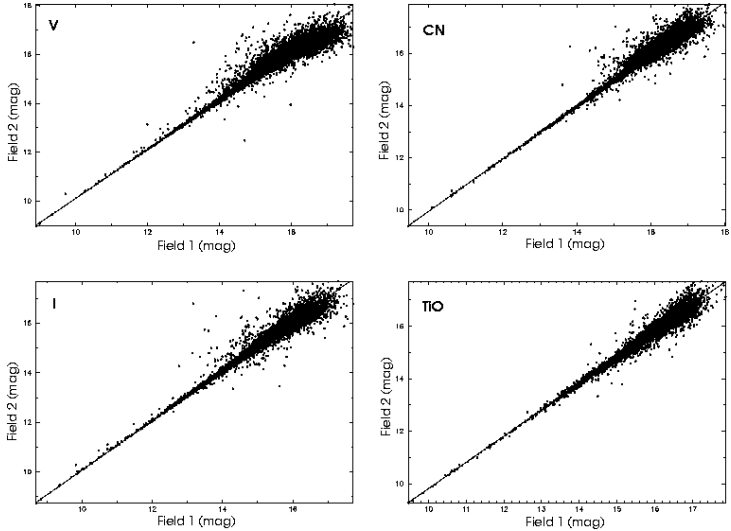}
\end{center}
\caption{The photometric offsets between the instrumental
magnitudes of Field 1 and Field 2 for each filter. Each field is
identified with a subscript.}
\label{fig:calF1F2}
\end{figure}

The master astrometric catalogue was corrected for zero-point
instrumental magnitude offsets between each observation, chip and
field.  These values were
calculated from the identification of common objects in the master
catalogue.  Figure \ref{fig:calF1F2} shows the calibration data for
stars common to Field 1 and Field 2 for each filter.  The
lack of scatter, other than the expected photometric errors, confirms
that the cross-identification of common objects works very well.  The
average error for all measured magitude offsets is approximately 0.01
magnitudes and the standard deviation from the fit for stars with an
instrumental magnitude greater than 14 is approximately 0.02
magnitudes.  This is also a measure of the quality of the
flat-fielding using stars common to Field 1 and Field 2.  If any
gradients exist then 
one would observe a systematic offset in Figure \ref{fig:calF1F2}
from flat-fielding errors.  One potential problem is that the
calibration of each chip is dependent only on adjacent chips, thus
the offset will inherit errors from every other chip other than the CCD
selected as the zero point reference.  A CCD chip that is 5 CCDs away
from the reference chip could suffer from a large (0.1 magnitude)
systematic offset.  This effect was monitored by plotting the colour
magnitude diagram for the reference chip overlaid with that for the CCD chip
being corrected.  Examination of the CMDs on opposite sides of the
mosaic shows no difference greater than 0.05 magnitudes,
which is the accuracy at which any offsets could be detected through
the examination of colour-magnitude diagrams.

\begin{figure}[htbp]
\begin{center}
\leavevmode
\includegraphics[width=7cm,height=6.41cm]{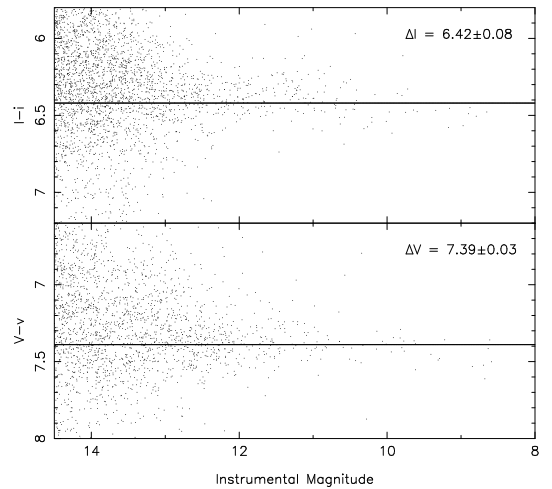}
\end{center}
\caption{A comparison of CFHT instrumental photometry with the DIRECT
project for the V and I filters.  Uppercase letters refer to standards
taken from the DIRECT project and lower case letter refer to the instrumental
photometry from this study.  Also shown is the zero point offset
adopted for each filter.}
\label{fig:cal1}
\end{figure}

With all of the photometry set to a common instrumental photometric
system, transformation to the standard system for the V and I filters
was computed using data from the DIRECT project \citep{mac01}.  
Comparison of the two photometry data sets is shown in
Figure \ref{fig:cal1}.  The scatter in this fit
for stars brighter than V, I=12 is 0.03 and 0.08
magnitudes respectively.  This is consistent with the internal
magnitude calibration of the DIRECT project, as seen in Figure 9 of
\citet{mac01}.  With errors this large, a reliable colour term
could not be determined.  Instead, only bright stars with a V$-$I colour
less than 0.3 magnitudes were used to determine the zero-point offsets;
otherwise the average colour terms for CFH12k from the 
online observer's manual were used.  The adopted transformation equations
are
\begin{equation}
\begin{array}{l}
V = V_{{\rm i}} + 7.39 + 0.001(V_{{\rm i}} - I_{{\rm i}}) \\ 
I = I_{{\rm i}} + 6.42 - 0.010(V_{{\rm i}} - I_{{\rm i}})
\end{array}
\end{equation}
where $V_{{\rm i}}$ and $I_{{\rm i}}$ are the observed instrumental magnitudes.

Calibration of the TiO and CN magnitudes was much easier.  It is
expected that the CN$-$TiO measurements for stars not on the AGB, such
as the MS should have CN$-$TiO$\simeq$0.  These stars lack the strong
TiO and CN absorption bands found in the cooler AGB stars and the TiO and CN
filters lie on no strong absorption features.  The TiO magnitudes
were adjusted such that CN$-$TiO has an average of zero for stars
with V$-$I less than 0.8.  The true magnitude of a star
observed in these filters is irrelevant, as only the difference between
them provides a measurement for identification of C and M-stars. 

\begin{figure}[htbp]
\begin{center}
\leavevmode
\includegraphics[width=7cm,height=6.77cm]{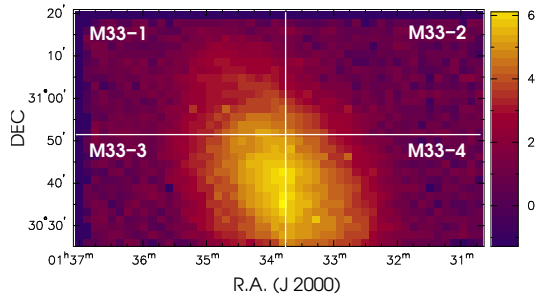}
\end{center}
\caption{Spatial map of all observed stars with completeness
corrections. The density scale is ln(number of stars per square
arcsec). The approximate boundaries of each target field is also shown.}
\label{fig:starmap}
\end{figure}

\subsection{Completeness Tests} \label{addstar}

Many aspects of this work involve relative counts of star types.
When observing a large extended object such as a galaxy, the detected
number of stars will vary from region to region due to properties of the
galaxy and constraints due to instrumentation.  In
order to compare relative statistics across the galaxy the
completeness needs to be known as a function of position.

Estimating detection limits due to the galaxy's structure requires
understanding knowledge of the poorly understood dust extinction.
Looking at Figure \ref{fig:m33V}, or any B or V band image 
of M33, it is 
easy to identify extinction in spiral arms that is caused by dust.
Star counts in dusty region will be lower as more stars will fall
below the detection limits.  

Detection limits due to instrumental constraints are primarily due to
a lack of resolution.  Detecting a star requires isolation of its PSF
on an image.  In very crowded fields it becomes impossible to separate
each stellar component. It is possible to test the confusion limit through
Monte-Carlo methods.

The Monte Carlo tests, now known as {\it addstar tests},
artificially add stars to a frame and attempt to recover them under the
same conditions as the original photometry was obtained.  Comparing
the input star list to the recovered objects gives a measurement of
detection limits and completeness due to instrumental
effects. No information is gained about extinction due to dust.  Addstar
tests were performed using the DAOPhot and ALLSTAR packages,
making use of the ADDSTAR routine.  This routine uses the model PSF
that was generated from the original photometry extraction to add
artificial stars to the image.  The star must be added with the same
noise characteristics as a star of the same instrumental magnitude.
If addstar tests are to be a valid estimation of the true
completeness, then the test must not significantly alter the
crowding statistics in the image 
when adding stars to it. The input stars must also have colour indices similar
to the original stars in the image.  Thus, the original photometry list
is used to generate the artificial input stars.  

We added 1000 stars per frame.  This number appeared not to make
even the most sparse fields (which contain just over
1000 stars) over-dense. Analysis of the sparse fields showed that
99.8\% of stars added at 100 sigma level were recovered, adding
confidence that our choice of 
adding 1000 stars per frame would not affect crowding statistics.

To generate the magnitude of an artifical star and its relative colour
indices, binned colour-magnitude 
diagrams and colour-colour diagrams, also called Hess diagrams, (see
\S\ref{cmd} and \S\ref{tcd}) are
used to determine the probability of generating stellar parameters.
First a CN$-$TiO versus V$-$I diagram is binned by 0.1 mag.  The
number of stars in each bin
is divided by the sum of all the bins to generate a
probability.  A random number 
between 0.000 and 1.000 is generated and bins are summed by row (V$-$I) then
incremented in column (CN$-$TiO) until the sum of the bin is greater than the
random number.  This bin determines the CN$-$TiO and V$-$I indices of the
artificial star.  Next a CN versus CN$-$TiO normalized grid is used to
determine the CN magnitude based on the corresponding CN$-$TiO column
and likewise a V vs V$-$I grid was used to determine the corresponding V
magnitude.  The artificial CMDs generated in this manner appeared
identical to the original data 
set.  The artificial magnitudes were then transformed back to the
instrumental magnitude system using the calibrations from
\S\ref{photcal}. The generated number of C-stars was close to 100, as   
there is approximately 1 C-star for every 100 other types of star.
Since C-star completeness tests is important in our analysis, the
artificially generated star list was changed by
taking the logarithmic value of each bin in the weighting grids before
normalization.  This places more weight on generating stars with small
populations.  Close to a hundred carbon stars were generated with this
alteration, giving good statistics for completeness determinations.

There were 51 exposures between
all four filters and there are 12 chips per exposure.  The addstar test was
run 10 times to obtain good statistics for the
completeness of the entire stellar population.  In total,
$6.12 \times 10^6$
stars were added to the frames.  Before stars were added to a frame, each
CCD for the same field of view was registered using MONTAGE2 from the
DAOPhot package to match star
co-ordinates, using the plate solutions from \S\ref{registration} to
aid in cross identification afterwards. 
Geometrically altering an image can introduce noise and artificial
artifacts, but is minimized if the image is oversampled.  The FWHM of a star should be greater than 2 pixels
to avoid registration artifacts. Under the best seeing conditions the
FWHM was just under 3 pixels in our images. For each image the
photometry steps from  
\S\ref{photometry} were repeated.

\begin{figure}[htbp]
\begin{center}
\leavevmode
\includegraphics[width=7cm,height=6.62cm]{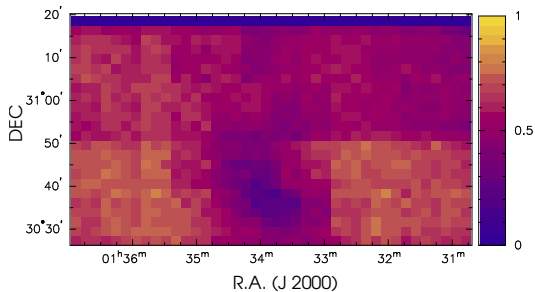}
\end{center}
\caption{This figure shows the relative
completeness for all observed regions.  Field 3 (bottom left) has
more observations than other regions, such as Field 2 (upper right)
and has a higher completeness level.} \label{fig:incompmap}
\end{figure}

The new photometry lists, containing the artificial stars, were
matched to the artificial star catalogue, and completeness statistics
were gathered.  In Figure \ref{fig:incompmap} the global
completeness level is plotted as a function of position. The spatial scale is
identical to Figure \ref{fig:m33V}.  One can see that each of the four
pointings has a different completeness level, which is due to changing
seeing conditions, and that Field 3 had more observations that any other
field and hence detected fainter stars.  

\section{Results}

\subsection{Star Counts}
 
In Figure \ref{fig:starmap}, the spatial distribution of stars appears
relatively uniform.  This is different from what is observed in star count
distributions for M31, as presented in Figures 2 and 3 of \citet{fer02}.  M31, like the Milky Way, has 
dwarf spherical companions, and their presence can be detected by
streams of stars sharing common orbits.  In the Milky Way there is the
Sagittarius Dwarf Galaxy, which is currently being sheared apart
through gravitational interaction with our Galaxy.  Mapping the
spatial distribution of evolved stars, such as C-stars or
RR Lyrae type variables \citep{viv02}, reveals tidal tails from
the Sagittarius Dwarf.  Streams of star have also been observed
emanating from globular
clusters such as Pal 5 \citep{ode01}.  Thus, if M33 has unseen
companions their presence could be betrayed in a complete star
map by selecting specific stellar populations. 

Visual examination of the raw star map for M33 does not reveal any
obvious perturbations.  However, if accreted satellites have no luminous 
stellar component (\citet{whi78} \citet{dek86}) their
presence could be hidden in stellar density maps.  A more detailed
exercise is to examine distributions of specific stellar  
populations that represent different epochs of star
formation, tracing the dynamical history of the galaxy through
perturbations of its stellar population.  To do this, we need to examine the
colour-magnitude diagrams of M33 and identify relevant populations.

\subsection{Colour-Magnitude Diagrams} \label{cmd}

\begin{figure}[htbp]
\begin{center}
\leavevmode
\includegraphics[width=7cm,height=7.44cm]{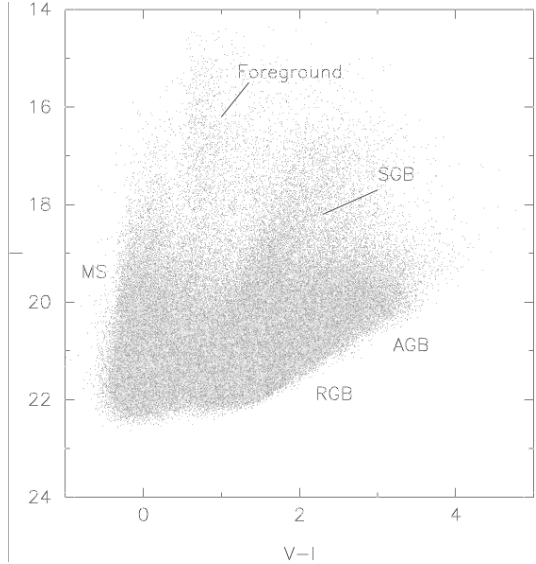}
\end{center}
\caption{I, V$-$I colour magnitude diagram for stars with V$-$I errors
less than 0.05 mag (choosen to allow various stellar populations to be
easily perceived). 
The major stellar populations are indicated as described in the text.}
\label{fig:cmdi}
\end{figure}

In Figure \ref{fig:cmdi}, a calibrated colour-magnitude
diagram (CMD) for I versus V$-$I is presented for
stellar objects from our survey with 
errors in the colour term less that 0.05 mag.  The MS, SGB, RGB and
AGB are all visible.  In Figure 
\ref{fig:cmdi} the MS is 
seen as a strong vertical band between $-$0.5 $\la$ V$-$I $\la$ 0.35.  This
represents young, luminous blue stars and provides a tracer of very recent
star formation.  Using Padova theoretical isochrones \citep{gir00} for a
young stellar population (age $6.31\times 10^{7}$ yr) and adopting
a distance modulus of 24.64 \citep{fre91} gives a mass of approximately
$5.8\ M_{\sun}$ at V=22 for a stellar population with Z=0.008. 
The large number of MS stars is no surprise, as the
spiral arms in M33 are regions of active star formation.

The red SGB is seen as a band of 
stars extending to V $\simeq$ 19, V$-$I $\simeq$ 2 out from the RGB
which is seen as a 
large clump centred at V $\simeq$ 22.5  and V$-$I $\simeq$ 1.5.
The AGB population, where C-stars will be found, 
is the band of stars with V$-$I $\ga$ 2  and V $\la$ 21. Foreground
contamination from Galactic stars  is seen as a
vertical sequence at V$-$I $\simeq$ 0.8 extending up to V=15,
the saturation limit of the detector.  This CMD is a very useful tool for
isolating specific stellar populations as will be seen in
\S\ref{cmdstars} and \S\ref{tcdstars}.

The distribution of stars on the CMDs is a result of star formation,
stellar evolution and extinction.  The effects of star formation and
stellar evolution are
observed through the presence of very young OB stars and older RGB and
AGB stars.  Extinction has the effect of blurring the CMD by shifting
observations dimmer and redward.  Since the amount of extinction 
depends upon the line of sight, stars with the same intrinsic
luminosity and colour can appear at different locations in the CMD.

\subsection{CMD Selected Star Counts} \label{cmdstars}

The substructure of halos and disks of nearby galaxies contains clues
about hierarchical galaxy formation.  State-of-the-art
simulations (\citet{kly99}, \citet{moo99}) show that accreted
sub-halos last much longer than previously thought, with the central
core lasting several tidal timescales \citep{hay02}, and
several hundred cores could reside in galaxies like the Milky
Way and M33.  

\begin{figure}[htbp]
\begin{center}
\leavevmode
\includegraphics[width=7cm,height=7.90cm]{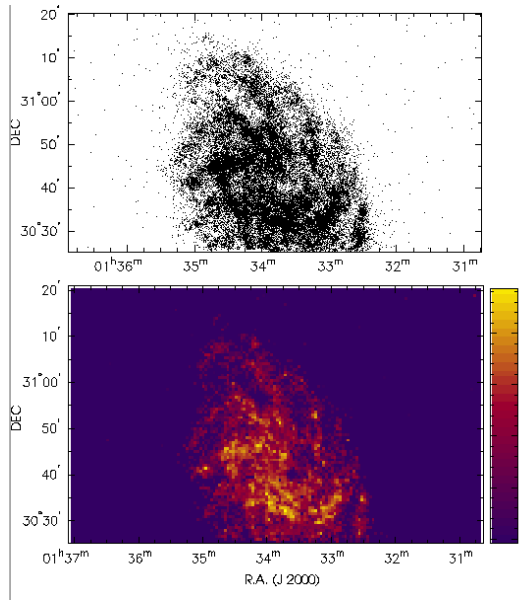}
\end{center}
\caption{The surface density
map of MS stars across the survey area. 
All stars with V$-$I 
$<$ 0.35 are plotted.  In the top panel the galaxy centre
appears to have few MS stars because of incompleteness due to extreme
stellar crowding. The bottom panel shows the binned surface density map for 15
$<$ V $<$ 22.  This map has been completeness corrected.  The
density scale is ln(number of stars per square arcsec).} 
\label{fig:msmap}
\end{figure}

Our survey data allows us to select specific stellar populations over
most of the M33's disk.  The observed MS reflects recent star
formation since the MS lifetimes for the massive and luminous stars is
short (under 1 Gyr).  MS stars were chosen as stars
with V$-$I colours less than 0.35 mag.  The resulting distribution is shown
in Figure \ref{fig:msmap}.  As should be expected, the spiral arm
patterns seen in Figure \ref{fig:m33V} are also well traced by
luminous MS stars.  The bottom panel of Figure
\ref{fig:msmap} is a binned map of MS stars with 15 $<$
V $<$ 22 that has been completeness corrected using
the data from \S\ref{addstar}.  This map shows all MS stars with masses
greater than $6 {\rm M}_{\sun}$ and the major spiral arms of the galaxy are
well traced.  The centre of M33 appears as a hole, as
stellar crowding is too great to allow reliable detection of any stars
in the region.  The application of stellarity cuts eliminated all
detections in this region of the galaxy.   

\begin{figure}[htbp]
\begin{center}
\leavevmode
\includegraphics[width=7cm,height=7.56cm]{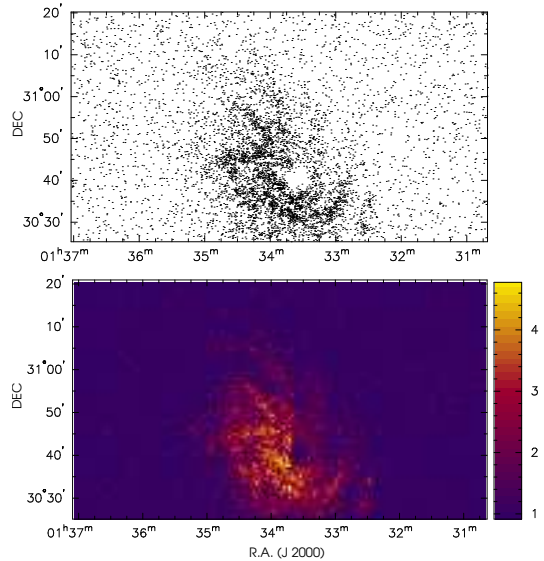}
\end{center}
\caption{Surface density map
of Super Giant stars across the survey area.  All stars with V$-$I $>$ 1.2
and V $<$ 21.2 are plotted.  The bottom panel shows the binned surface
density map and has been completeness corrected.  The density scale
is ln(number of stars per square arcsec).}
\label{fig:sgmap}
\end{figure}

The same
exercise can be applied to the SGB population.  SGB stars are identified
by selecting all stars with a V$-$I colour greater than 1.2 and a
V magnitude less than 21.75.  Faint stars are excluded to avoid RGB stars
at the base of the RGB clump.  The distributions of SGB stars is shown in
Figure \ref{fig:sgmap}.  Like the MS stars,
the SGB population is 
relatively young and will trace out stellar populations with ages less
than about 1 Gyr.  The SGB map suffers from foreground contamination by
M-dwarfs, seen as
a scatter of stars over all observed fields, but the galaxy is still
identifiable.  The SGB population is largest in areas containing the
most MS stars.  This occurs because both groups of stars have similar ages
and the MS lifetimes are longer than stars found in the SGB phase.
Both the SGB and MS maps (Figures \ref{fig:msmap} and \ref{fig:sgmap})
trace out a high density structure that almost 
encloses the centre of the galaxy.  This feature starts on the east
side of the galaxy extending south of the galaxy's centre and then to
the west of the galaxy where it then sharply turns eastward toward the
galaxy centre and then blends into two spiral arms that extend
northward. 

\begin{figure}[htbp]
\begin{center}
\leavevmode
\includegraphics[width=7cm,height=7.52cm]{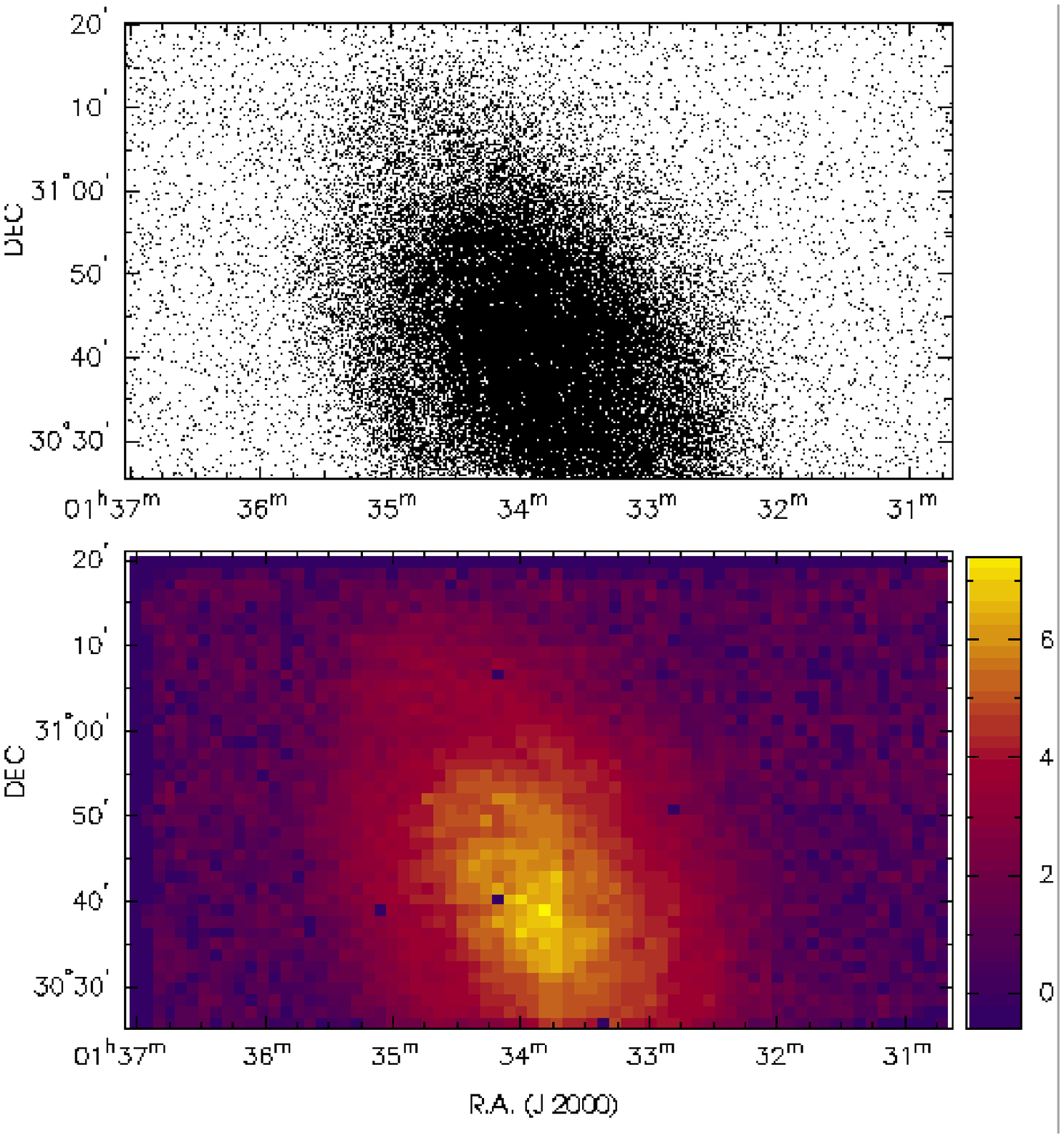}
\end{center}
\caption{Surface density map of AGB
stars across the survey area.  All stars with V$-$I $>$ 2.0 and 19 $<$ I $<$
21 are plotted.  There are instrumental artifacts towards the center
of the galaxy due to incompleteness.  The lower panel shows the binned
surface density map.  This map has been completeness
corrected.  The density scale is ln(number of stars per
square arcsec).}
\label{fig:agbmap}
\end{figure}

The spatial distribution of AGB stars is shown in Figure 
\ref{fig:agbmap}.  The population shows a
smooth distribution of stars 
compared to the clumpy distribution of MS stars.  This is due to
dispersion of the AGB population.
M33 itself shows many localized regions of massive star birth,
such as the gas complex NGC 604.  Over time, new stellar
associations will disperse.  Thus, the distribution of MS stars will
appear clumpier than that of AGB stars.  

The AGB star map does not show any extended structure apart from
gently tracing out the spiral arms.  At semi-major radii larger than
30\arcmin, the AGB population of M33 is lost due to foreground
confusion from Galactic
red dwarfs.  Just as with MS stars, incompleteness is
strongest towards the centre of M33, creating an apparent hole.

\subsection{Colour-Colour Diagrams}\label{tcd}

\begin{figure}[htbp]
\begin{center}
\leavevmode
\includegraphics[width=7cm,height=7.33cm]{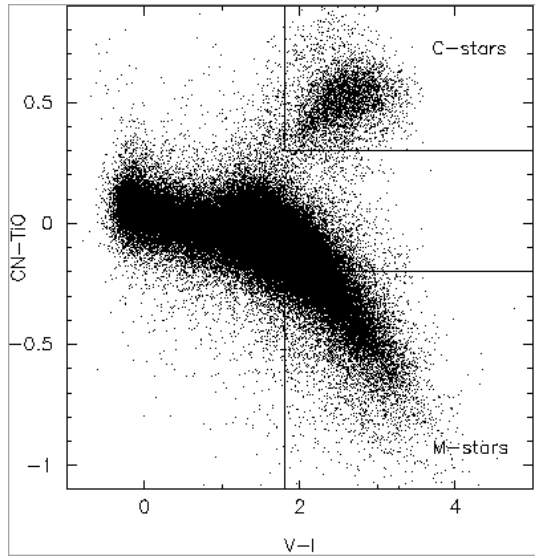}
\end{center}
\caption{Colour-colour magnitude diagram for
CN$-$TiO versus V$-$I for all stars with errors less than 0.05 magnitudes
in CN$-$TiO and V$-$I.  The C-stars and M-stars are clearly
differentiated in this diagram.  The few stars with V$-$I > 1.8 and
between the C and M-star region are likely S-stars.}
\label{fig:ccd}
\end{figure}

The broad-band photometry can be combined with the narrow-band
photometry to  discriminate between spectral sub-types, namely C-stars and
M-stars.  Figure \ref{fig:ccd} shows the colour-colour diagram for the
entire M33 field.  The carbon stars, which have strong CN absorption
bands appear as an isolated group with an average CN-TiO index of
0.5.  The M-stars, with strong TiO absorption bands, tail-off from the
RGB population at approximately V$-$I=2 towards redder colours or
later spectral type. 
There is a difference of about 1 magnitude between the CN-TiO index
values of C-stars and M-stars; the CN and TiO filters clearly do a
good job at selecting AGB sub-types. 

M31 has a distance modulus of 24.47 \citep{dur01} and M33 has
a measured distance modulus of 24.64 \citep{fre91}. The 1$\sigma$ errors in the
measurements are approximately $\pm$ 0.15, thus, to within 1$\sigma$
M31 and M33 are at the same distance.  To define the selection boxes for 
choosing M-stars and C-stars, the criteria of \citet{bre96} for M31 are
adopted.  C-stars are identified 
with CN$-$TiO $>$ 0.3 and M-stars with CN$-$TiO $< -$0.2 and both types must
have V$-$I $>$ 1.8.  These were originally chosen by
spectrally identifying C and M-stars and selecting colour values that
encompassed C-stars without contamination (see \citet{bre96} for
details).  

\begin{figure}[htbp]
\begin{center}
\leavevmode
\includegraphics[width=7cm,height=7.29cm]{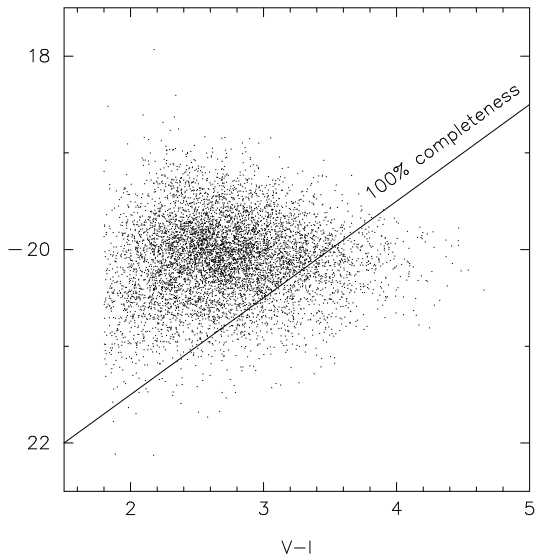}
\end{center}
\caption{Colour-magnitude
diagram for all C-stars.  The 100\% detection completeness limit is
shown.}
\label{fig:cmdcstar}
\end{figure}

Figure \ref{fig:cmdcstar} shows a colour-magnitude diagram for the 7936
C-stars selected using the adopted criteria.  The completeness limit
will be discussed later in this section.  As expected, these stars
occupy the AGB branch location of the CMD.  The foreground
contamination of C-stars from the Milky Way is unimportant, 
the surface density being only $0.019\ {\rm deg}^{-2}$ \citep{gre92}
down to V=18. 

\begin{figure}[htbp]
\begin{center}
\leavevmode
\includegraphics[width=7cm,height=6.32cm]{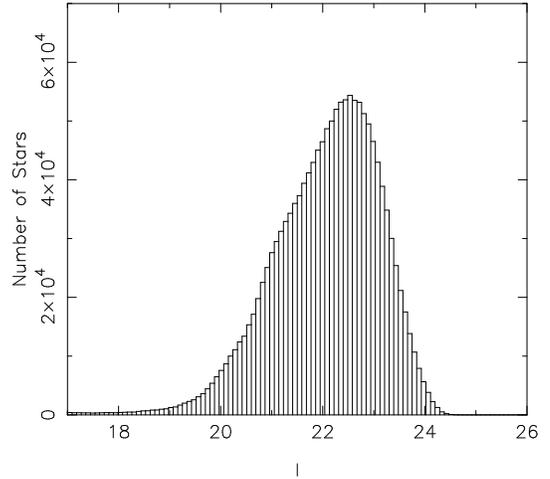}
\end{center}
\caption{Luminosity
function for all detected objects.  The turnover at 22.5 mag gives an
estimation of the
completeness limit.}
\label{fig:completeI}
\end{figure}

The M-star population will suffer strong
contamination from the Milky Way.  NGC 6822 (at b=$-$18\arcdeg) was
observed to have approximately 9000 foreground stars over a
28\arcmin$\times$42\arcmin \ field of view \citep{let02}.  M33
(b=$-$31\arcdeg), while at a higher Galactic latitude, will still
suffer a substantial  M-star foreground contamination (see
\S\ref{tcdstars}).  Contamination from non-AGB members within M33 is
not a serious  problem, as members of the RGB and SGB populations will not have
strong CN or TiO absorption bands.  However, choosing an I-band
magnitude cut will limit foreground contamination and contamination
from non-AGB stars in M33.  Using Figure \ref{fig:cmdcstar}, C-stars
and M-stars also have I magnitudes between 18.5 and 21.  These values
were chosen to enclose a majority of the detected C-stars, with out
straying far below the 100\% completeness limit.  To quickly estimate
the completeness limit in the I-band, the raw luminosity function, shown in
Figure \ref{fig:completeI}, was used.  The number of stars rises
approximately  linearly towards fainter magnitudes, until
approximately I=22 mag, when the number of objects quickly declines as
the detection limit is reached.   

\subsection{Colour-Colour Diagram Selected Star Counts} \label{tcdstars}

\begin{figure}[htbp]
\begin{center}
\leavevmode
\includegraphics[width=7cm,height=4.21cm]{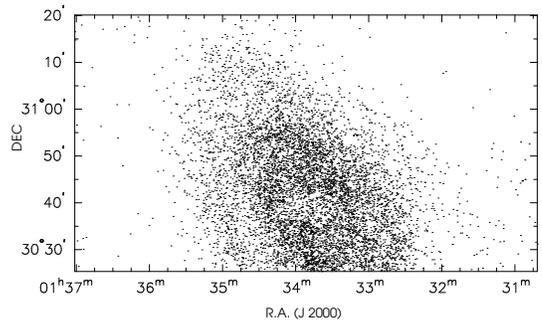}
\end{center}
\caption{Spatial map of C-stars.  As
seen in previous star-maps, incompleteness is strongest towards the
centre of the galaxy giving the appearance of a hole.}
\label{fig:csmap}
\end{figure}

As a continuation of \S\ref{cmdstars} we can now examine the AGB
stellar content of M33.  Figure \ref{fig:csmap} and the top panel of Figure
\ref{fig:csmapavg} show the spatial distribution of C-stars in M33.
The distribution of C-stars traces out the extent of M33's disk
well, as there is very little foreground contamination.  Visual
comparison of Figure 
\ref{fig:csmap} with the MS map in Figure \ref{fig:msmap} shows that
the C-star distribution does exhibit some spiral structure.  This is
not unexpected, as the C-star population represents intermediate aged
stars that 
were produced in spiral arms just like the current population of young
stars.  Their velocity dispersion as well as differential galactic
rotation has slowly begun to smooth out the older
population. 

\subsubsection{Tidal Interactions} \label{tidal}

\begin{figure}[htbp]
\begin{center}
\leavevmode
\includegraphics[width=7cm,height=10.74cm]{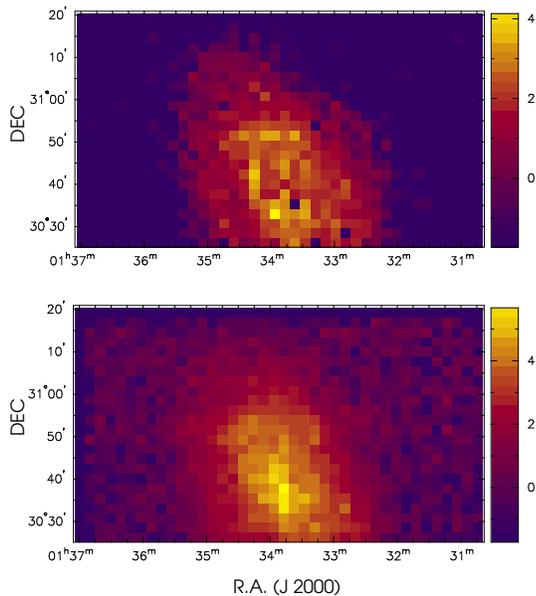}
\end{center}
\caption{The top panel shows the binned surface
density map of C-stars and the lower panel shows the M-stars. Both
maps have been completeness corrected.  The density scale is
ln(number of stars per square arcsec).}
\label{fig:csmapavg}
\end{figure}

There is no evidence of tidal disruption within the C-star population, neither
is there indication of C-stars originating from a different
system. This is in contrast 
to the claim of \citet{blo04} who suggest that M33 displays
``spectacular arcs of carbon stars'' beyond 14\arcmin \ from its
nucleus.  The lack of
external interactions may be a reason why M33 displays beautiful {\it
grand design} spiral arms that can be traced from the outer most
regions of the disk directly towards the center of the galaxy.  The
bottom panel in Figure \ref{fig:csmapavg} shows the corresponding
M-star distribution.  It is similar to the C-star distribution, except that
foreground contamination, from Milky Way M-dwarfs, is stronger.  

\subsubsection{Radial Distributions}

\begin{figure}[htbp]
\begin{center}
\leavevmode
\includegraphics[width=7cm,height=7.25cm]{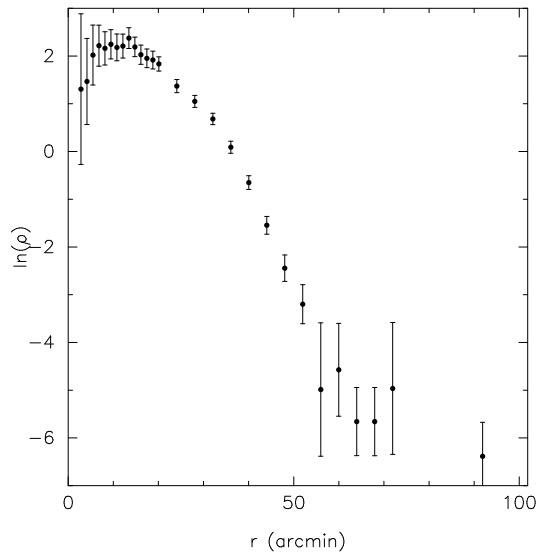}
\end{center}
\caption{Stellar density profile for C-stars
in M33.  The units of density are ln(${\rm number\ of\
stars\ per\ arcmin}^{2}$).}
\label{fig:cstarpro}
\end{figure}

\begin{figure}[htbp]
\begin{center}
\leavevmode
\includegraphics[width=7cm,height=7.34cm]{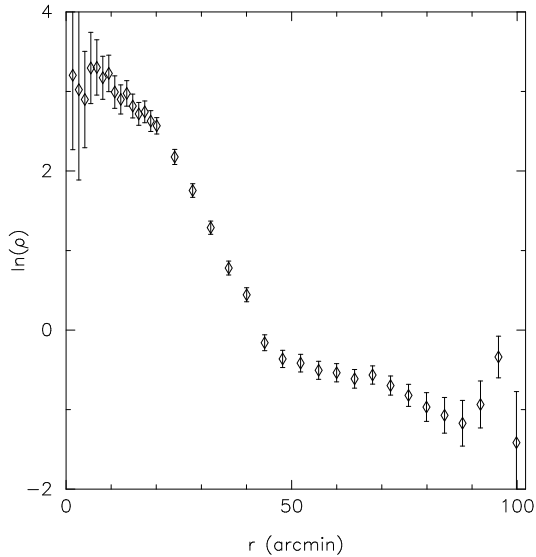}
\end{center}
\caption{Stellar density profile for M-stars
in M33.  The units of density are ln(${\rm number\ of\
stars\ per\ arcmin}^{2}$).}
\label{fig:mstarpro}
\end{figure}

The radial distribution of stars in a galaxy allows a quantitative
measurement of the galaxy's morphology.  To extract radial profiles
from M33, its tilt must be taken into account.  To do this, shape
parameters for M33 from the Third Reference Catalogue of Bright Galaxies
(RC3) were obtained from the NASA Extragalactic Database (NED),
specifically
the length of the semi-major axis and the ratio of the semi-major axis
to the semi-minor axis.  If M33 was seen face on its shape
would be a circle.  Ellipses centred on M33 were constructed with
different radii, and star counts were made for each radius using
completeness corrected C and M-star counts.  Figures
\ref{fig:cstarpro} and \ref{fig:mstarpro} show the deprojected radial
profile for M33 for C-stars and M-stars, respectively, in units of
ln(${\rm number\ of\ stars\ per\ arcmin}^{2}$).  

Examining the C-star profile, we see that it is flat from the centre
of M33 out to 15\arcmin \ and then the number density of stars decreases
out to 50\arcmin.  Here the slope changes again, and becomes flat
out to approximately 70\arcmin \ beyond which there are too few C-stars to
provide useful statistics.  The M-star profile is qualitatively
similar to the C-star profile.  The number density of M-stars
decreases out to
about 20\arcmin \ from the centre of the galaxy, where there is a steepening of the slope and the distribution drops off.  This feature can be seen in
the lower panel of Figure \ref{fig:csmapavg} as a separation between the inner and
outer disk in the  distribution.  At 45\arcmin, the
foreground population of M-stars becomes dominant, reducing the slope
of the M-star population out to the edge of the field of view.

\subsubsection{Metallicity Gradients}

\begin{figure}[htbp]
\begin{center}
\leavevmode
\includegraphics[width=7cm,height=7.46cm]{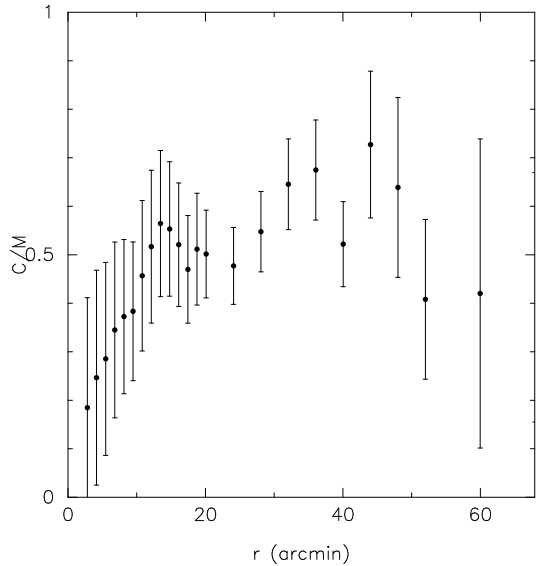}
\end{center}
\caption{The C/M-star ratio as a function of
galactocentric radius (note: 1 arcmin $\simeq\ $ 0.25 kpc at a distance
modulus of 24.64).}
\label{fig:cmstarpro}
\end{figure}

The time needed for a stellar population to produce the majority of its
carbon stars is about 1 Gyr, thus for stellar populations older than
this the C/M-ratio is independent of the star formation
history \citep{mou03}. 
Since the ratio of C-stars to M-stars is a tracer of metallicity, the
C-star and M-star profiles can be converted into a completeness
corrected C/M-ratio profile.  Before this can be done, the foreground
population of M-stars needs to be estimated.  It is assumed that this
population of M-stars is uniform across the field. The M-star
profile is then used to estimate the foreground population.  This was done
by assuming that the M-star profile has a constant value 
beyond 25\arcmin \ obeying an exponential disk profile.  The foreground
M-star population was in this way estimated to be $0.50\pm0.03\ {\rm
arcmin}^{-2}$.  
This gives approximately 2300 foreground M-stars over the field of view.  As a
check the foreground population can also be easily calculated by
counting stars at the periphery of the image.  All M-stars with an RA
greater than 1$^h$ 36$^m$ were considered to be foreground M-stars as this
area contains few C-stars.  The completeness corrected stellar
density of M-stars in this region was found to be $0.55\pm0.03\ {\rm
arcmin}^{-2}$ 
consistent with our derived value.  Figure \ref{fig:cmstarpro} shows
the C/M-ratio as a function of galactocentric radius.  The ratio
increases to a radius of 12\arcmin \ and then flattens for the outer
disk regions.  This result indicates that the metallicity of
M33 is high in the centre and low in the outer
parts of the disk, with a change in the gradient along the way.  This
is compatable with other metallicity gradient measurements \citep{vil88}.

These results are consistent with viscous disk formation
models that predict exponential surface luminosity
profiles of spiral galaxy disks \citep{zar92}.  The rotation curves of spiral
galaxies show solid body rotation in the inner parts of the disk and
flat rotation curves in the outer part of the disk, where the rotation
curve is dominated by dark matter.  For solid body rotation there is
no angular velocity difference between material at different radii, meaning
that there is no viscous drag or turbulent diffusion.  In the outer
parts of the disk, the opposite is true with the production of radial
gas flows.  In galaxy formation models, negative abundance gradients
are produced \citep{som90}, the evolutionary effect of rotation then
smooths the metallicity gradient in the outer disk where the rotation
curve is flat.  Radial outflows will transfer metal-rich material into
metal poor material, and vice versa with radial inflows.  This produces
a metallicity distribution with a change in slope where the rotation
curve flattens.  Examination of the 21cm rotation curve for
M33 \citep{cor00} reveals that, in fact, the rotation curve flattens
around 10$-$15\arcmin, consistent with our findings.

\begin{figure}[htbp]
\begin{center}
\leavevmode
\includegraphics[width=7cm,height=13.98cm]{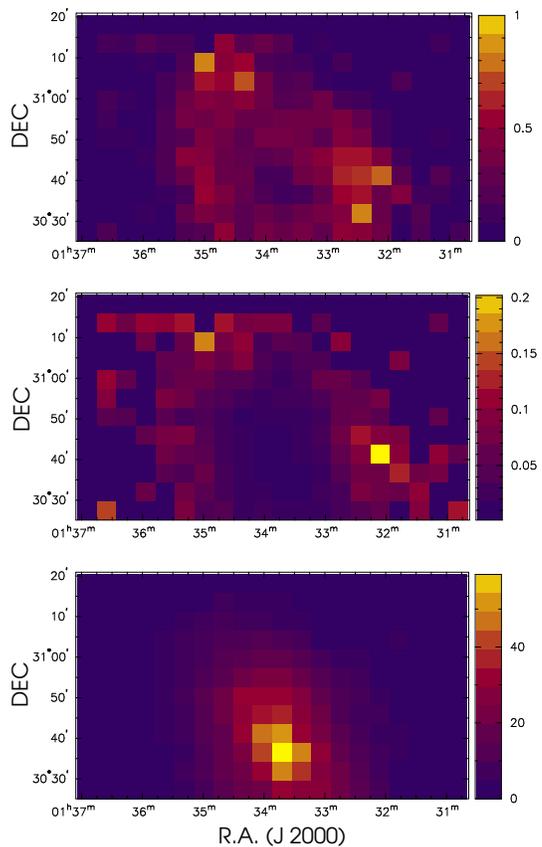}
\end{center}
\caption{The top panel shows the spatial map of the C/M-star
ratio.  The middle panel
shows the associated error in the C/M-star ratio and the bottom panel
plots the signal-to-noise ratio.}
\label{fig:cmmap}
\end{figure}

The current data set allow for a C/M-ratio map for M33.
This completeness corrected map is shown in Figure \ref{fig:cmmap}.
The increase in the C/M-ratio with increasing radius is apparent, as are two
regions with a high C/M-ratio.  The areas of enhanced C/M ratio are
located in the regions  where the arcs mentioned in Block et
al. (2004) are to be found. The number of M-stars and
C-stars drops towards the edge of the disk, thus the error in the
C/M-star measurement will be higher in these regions.  The second and
third panel in Figure
\ref{fig:cmmap} show the associated error
and the signal-to-noise ratio (S/N) for each region in the C/M-ratio map.
The peak in the northern region of high C/M has a S/N less than 3 and
its associated error in the ratio is approximately 0.35, making the
detection of the peak with the bin size used somewhat uncertain.  The same
argument applies to 
the south-west region of the map.   

\begin{figure}[htbp]
\begin{center}
\leavevmode
\includegraphics[width=7cm,height=7.44cm]{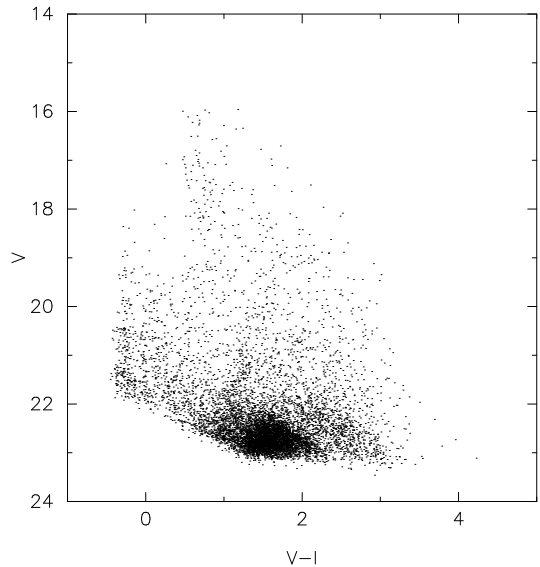}
\end{center}
\caption{Colour-magnitude diagram for V versus V$-$I for a 10\arcmin \ box centered at 
approximately 1$\rm ^h$35.45$\rm ^m$ +31$\arcdeg$10\arcmin, \ near the
edge of M33's visible disk.}
\label{fig:extract}
\end{figure}

To test whether the enhanced C/M-ratio regions
are statistically significant, the S/N can be increased by increasing
the bin size 
to avoid small number statistics.  The C/M-ratio map was rebinned with
an area four times greater.  The north and southwest regions each have
average C/M-ratios of 0.5 and 0.6 $\pm$ 0.04 respectively.  Other
regions around the edge of the disk show C/M-ratios no higher than
0.4. It thus appears that the higher C/M-ratios are real.  These
regions could simply mark the outer reaches of spiral arms with lower
metallicities.  Both regions do have spiral arm structure within them.  
Figure \ref{fig:extract} shows a CMD for a region located at the edge
of the visible disk of M33.  The stellar populations and features of
the CMD in Figure
\ref{fig:cmdi} are still visible.  However, the CMD
does not show any significant morphological differences.  A deeper survey is 
necessary to search for a distinctive stellar population in the region, that
could consist of an old, low-luminousity stellar population.

The first panel in Figure \ref{fig:cmmap} also shows that
the C/M-ratio is a function of galactocentric radius.
This gives the galaxy the appearance of being surrounded by a ring of
material with a lower metallicity.  It has recently been suggested
that the Milky Way is also surrounded by a ring traced by stars
of lower metallicitiy \citep{iba03}.  As suggested by Ibata et al. (2003),
this feature could be extended spiral arm structure.  In galaxy formation
simulations the edge of the disk is expected to be young and contain
metal poor gas \citep{nav97}.  In M33, the regions of low metallicity
also correspond to the edge of the disk, as traced by C-star and
MS-star populations.  In the Milky Way, distinct spiral arm structure has been
identified from 21-cm emission (\citet{dav72}, \citet{iba03}).  This similarity
suggests that the Milky Way ``ring'' may be consistent with spiral arms
and does not require tidal interaction of dwarf galaxies for its formation. We
may be seeing the same effect here in M33.

\subsection{C-star Luminosity Function} \label{lum}

\begin{figure}[htbp]
\begin{center}
\leavevmode
\includegraphics[width=7cm,height=19.35cm]{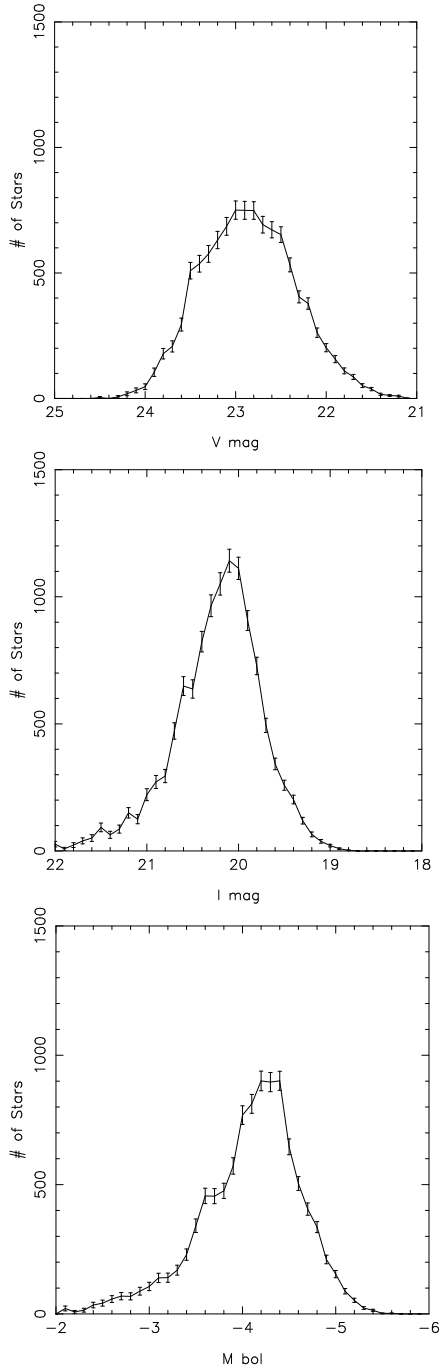}
\end{center}
\caption{Magnitude distribution
of C-stars for V-filter, I-filter and bolometric magnitudes.}
\label{fig:lum}
\end{figure}

This study has identified 7936 C-stars and Figure \ref{fig:lum}
show the corresponding completeness
corrected luminosity functions (LF) for V, I and bolometric
magnitudes. The LF is similar to those observed in other systems such
as M31 and the  SMC as
shown in \citet{gro02}.
To calculate absolute bolometric magnitudes for the C-stars we used a
distance modulus of 24.64 \citep{fre91} 
with the bolometric correction (BC) given by \citet{bes84} for
M-stars. 

\begin{equation}
\begin{array}{l}
M_{bol}=I + BC - 24.64 \\   
BC=0.3+0.38(V-I)-0.14 (V-I)^2.
\end{array}
\end{equation}

The C-star luminosity function (CSLF) has a narrow peak and thus has
potential as a good distance indicator 
(\citet{ric89}, \citet{gro02}).  The problem is disentangling the dependence
of the peak magnitude of the CSLF on galactic properties such as metallicity or
star formation history.  Figure 8 of \citet{gro02} shows that
the mean of the CSLF does not depend on [Fe/H], with
most systems having a mean bolometric magnitude between -4 and
-5.  Discrepancies can be explained either through incompleteness of
the CSLF at faint magnitudes causing the mean to be too bright, or the
absence of an intermediate age population leaving only faint C-stars.
In M33 the average bolometric magnitude is found 
to be $-$4.2 mag $\pm$ 0.1 which is similar
to M31 and the SMC (both have ${\rm M_{bol}}= -4.3$). Because such diverse
systems have similar C-star LFs, 
C-stars could be used as distance indicators.

\section{Conclusions}

Using a four-band photometric system, the AGB stars in the
nearby spiral galaxy M33 were classified into C and M-star types.  The
photometry catalogue allowed an examination of the different stellar
populations of M33.  M33 has a large number of MS OB stars being
produced by current star formation.   The extent of the disk and spiral arm
structure was shown through examination of the spatial distributions
of MS and SGB-stars.  The AGB population, being older, is dispersed
and only tenuously trails the spiral arm structure The distribution of
the AGB stars revealed no smaller galactic 
companions, such as those found in the local environments of M31 or
the Milky Way.   

Using colour-colour diagrams, the C-star
and M-star populations were used to map the C/M-star ratio.  The
C/M-star ratio is known to trace metallicity, and the C/M-ratio
profile and C/M-star map were produced.  The C/M-star profile shows a
metallicity gradient dependent on galactocentric radius.  The profile
was found to flatten at the same radius at which the radial-velocity
profile also flattens.  These results are consistent with viscous-disk
formation models where the metallicity gradient becomes flattened in
the outer part of the disk as material originating at different
initial radii become mixed.

The C/M-star map shows the outer parts of M33's galactic disk to be
metal poor.  This can give the appearance to an observer located
inside the galaxy that they are surrounded by a ring of metal poor
material.  The C/M-star map also shows two regions with an enhanced C/M-star
ratio.  These regions may be a natural occurrence at the end of a
spiral arm, or may trace a different underlying population.  These
regions will require deep follow-up photometric surveys to examine the
stellar populations in order to explain the implied lower metallicity in these
regions. 

\acknowledgements

This research has made use of the NASA/IPAC Extragalactic Database
(NED) which is operated by the Jet Propulsion Laboratory, California
Institute of Technology, under contract with the National Aeronautics
and Space Administration.

\onecolumn{

\begin{table}[htbp]
\begin{center}
\caption[List of Filters.]{Filters used in the Four Band Photometry
System}\label{ta:filters}
\ \\
\begin{tabular}{lccc}
\hline \hline
Filter & Central Wavelength (nm) & Bandwidth (nm) & Max. Trans. (\%)\\ \hline
Mould V & 537.4 & 97.4 & 94 \\
Mould I & 822.3 & 216.4 & 91 \\ 
TiO     & 777.7 & 18.4 & 92 \\
CN      & 812.0 & 16.1 & 95\\ \hline
\end{tabular}
\end{center}
\end{table}

\begin{table}[htbp]
\begin{center}
\caption[Coordinate of M33 Target fields..]{Target list}\label{ta:target}
\ \\
\begin{tabular}{lcc}
\hline \hline
Field ID & R.A. & DEC \\ \hline
M33-1 & $1^{{\rm h}}35^{{\rm m}}24^{{\rm s}}$ & $+31\degr 06' 30''$ \\
M33-2 & $1^{{\rm h}}32^{{\rm m}}18^{{\rm s}}$ & $+31\degr 06' 30''$ \\ 
M33-3 & $1^{{\rm h}}35^{{\rm m}}23^{{\rm s}}$ & $+30\degr 39' 30''$ \\
M33-4 & $1^{{\rm h}}32^{{\rm m}}18^{{\rm s}}$ & $+30\degr 39' 30''$ \\ \hline
\end{tabular}
\end{center}
\end{table}

\begin{table}[htbp]
\begin{center}
\caption[Observation Log.]{Observation Log}\label{ta:obslog}
\ \\
\begin{tabular}{llclcc}
\hline \hline
Date & Target & Filter & Exp. (s) & FWHM ($\arcsec$) & Airmass \\ \hline
Oct 30, 1999 & M33-3 & V   & 3 $\times$ 400 & 0.7, 0.7, 0.7 & 1.20,
1.18, 1.16 \\
             & M33-3 & I   & 3 $\times$ 200 & 0.6, 0.6, 0.6 & 1.10,
1.09, 1.08 \\ 
             & M33-4 & I   & 3 $\times$ 200 & 0.6, 0.6, 0.6 & 1.07,
1.06, 1.06 \\ 
             & M33-4 & V   & 3 $\times$ 400 & 0.7, 0.7, 0.7 & 1.05,
1.05, 1.04 \\ 
             & M33-1 & V   & 3 $\times$ 400 & 0.6, 0.6, 0.6 & 1.03,
1.03, 1.03 \\ 
	     & M33-1 & I   & 3 $\times$ 200 & 0.8, 0.8, 0.8 & 1.02,
1.02, 1.02 \\
	     & M33-2 & I   & 3 $\times$ 200 & 0.8, 0.8, 0.8 & 1.02,
1.02, 1.02 \\ 
             & M33-2 & V   & 3 $\times$ 400 & 0.9, 0.9, 0.9 & 1.03,
1.03, 1.04 \\
             & Sky Flat & V & 30, 21, 15, 11, 6, 5, 4 & -- & -- \\ 
Oct 31, 1999 & Sky Flat & I & 4, 5, 7, 9, 12, 17, 24  & -- & -- \\ 
	     & Bias & --   & 4 $\times$ 0    & -- & -- \\  
Dec 3, 2000  & M33-3 & TiO & 3 $\times$ 1000 & 0.6, 0.6, 0.6 &
1.07, 1.05, 1.04 \\ 
             & M33-3 & CN  & 3 $\times$ 1000 & 0.8, 0.8, 0.8 & 1.02,
1.02, 1.02 \\ 
	     & M33-4 & CN  & 3 $\times$ 1000 & 0.5, 0.5, 0.6 &
1.04, 1.05, 1.08 \\ 
             & M33-4 & TiO & 3 $\times$ 1000 & 0.7, 0.7, 0.9 &
1.12, 1.16, 1.21 \\ 
             & M33-1 & TiO & 3 $\times$ 1200 & 1.0, 1.3, 1,3 & 1.29,
1.38, 1.49 \\ 
             & Sky Flat & TiO & 3, 7, 4 & -- & -- \\ 
	     & Bias  & --  & 0        & --  & -- \\ 
Dec 4, 2000  & Sky Flat & CN  & 6 $\times$ 5,7 & -- & -- \\ 
             & Sky Flat & TiO & 2 $\times$ 12, 2 x 20, 30    & -- & --
\\
             & Sky Flat & CN  & 40, 60, 100, 140, 200 & -- & -- \\ 
             & Bias     & --  & 0     & --  & -- \\  
             & M33-1 & CN  & 3 $\times$ 1000 & 0.7, 0.6, 0.7 &
1.04, 1.06, 1.08 \\ 
             & M33-1 & TiO & 3 $\times$ 1000 & 0.7, 0.7, 0.7 &
1.13, 1.17, 1.22 \\ 
             & M33-2 & TiO & 3 $\times$ 1000 & 0.7, 0.8, 0.8 &
1.33, 1.42, 1.52 \\ 
             & M33-2 & CN  & 3 $\times$ 1000 & 0.8, 0.8, 0.8 & 1.72,
1.90, 2.12 \\
	     & Bias  & --  & 6 $\times$ 0    & --  & -- \\ \hline
\end{tabular}
\end{center}
\end{table}

\begin{table}[htbp]
\begin{center}
\caption[Object Catalog for M33.]{Object Catalog for M33
(abridged)}\label{ta:master} 
\ \\
\tiny
\begin{tabular}{lcccccccccccc}
\hline \hline
Star ID & R.A. & DEC & V & $\sigma_V$ & I & $\sigma_I$ & CN &
$\sigma_{CN}$ & TiO & $\sigma_{TiO}$ & $\chi$ & Sharp \\ \hline
      1 & 1 31 04.68 & 30 52 07.70 & 15.8437 &  0.0423 & 15.1750 &
0.0118 &  9.4495 &  0.0079 &  9.5432 &  0.0088 &  2.8480 &  0.0670 \\
      2 & 1 30 45.71 & 30 45 01.00 & 16.7190 &  0.0231 & 99.9990 &
9.9990 & 10.0849 &  0.0366 & 10.3589 &  0.0140 &  4.7347 &  0.1315 \\
      3 & 1 30 45.15 & 30 51 37.80 & 16.5790 &  0.0048 & 16.0423 &  
0.0095 & 10.3286 &  0.0040 & 10.2939 &  0.0036 &  1.4807 &  0.1440 \\
      4 & 1 31 03.82 & 30 51 36.90 & 17.4459 &  0.0576 & 99.9990 &  
9.9990 & 10.6256 &  0.1168 & 10.3792 &  0.1184 & 14.6062 &  0.9263 \\
      5 & 1 31 07.73 & 30 49 09.80 & 17.1538 &  0.1146 & 16.0570 &  
0.0768 & 10.2046 &  0.1594 & 10.2802 &  0.0882 & 18.9920 &  2.0965 \\ \hline
\end{tabular}
\normalfont
\end{center}
\end{table}

\clearpage
\centerline{\bf{TABLE CAPTIONS}}

{\sc Table} 1. {\sc Filters: Column (1) names the filter, column (2)
gives the central wavelength in nanometres, column (3) gives the
filter width in nanometres and column (4) gives the maximum
transmission as a percentage.}

{\sc Table} 2. {\sc Target List: Column (1) gives the field ID, columns
(2) and (3) gives the right accension and declination in J2000 co-ordinates.}

{\sc Table} 3. {\sc Observation log: Column (1) gives the
date of the observation, column(2) gives the target field
or type of calibration image, column (3)
gives the filter used, column (4) gives the number of exposures and
exposure time in seconds and column (5) gives the average
full-width-half-maximum (FWHM) of stars on each frame in arc seconds.}

{\sc Table} 4. {\sc Object catalogue: column (1) gives the star ID,
column (2) and column (3) give 
the right accension and declination in J2000 co-ordinates, columns
(4), (6), (8), (10) give the V, I, CN and TiO magitudes and columns
(5), (7), (9), (11) give the associated photometric error returned by
ALLSTAR.  Columns (12) and (13) give the Chi-squared and Sharp values
for the PSF fit from DAOPhot.}

}

\end{document}